\documentclass[10pt]{article}

\usepackage{amsmath}
\usepackage{amssymb}

\usepackage{graphicx}

\usepackage{cite}

\usepackage{color} 

\usepackage[labelfont=bf,labelsep=period,justification=raggedright]{caption}

\makeatletter
\renewcommand{\@biblabel}[1]{\quad#1.}
\makeatother

\date{}

\begin{document}

\begin{flushleft}
{\Large
\textbf{Quantifying the interplay between environmental and social effects on aggregated-fish dynamics}
}
\\
 \vspace{2 mm}
Manuela Capello$^{1\ast}$, 
Marc Soria$^{1}$, 
Pascal Cotel$^{1}$,
Jean-Louis Deneubourg$^{2}$,
Laurent Dagorn$^{3}$
\vspace{2 mm}
\\
\bf{1} UMR EME, Institut de Recherche pour le D\'eveloppement (IRD), Saint Denis, La R\'eunion, France 
\\
\bf{2} USE, Unit of Social Ecology, Universit\'e Libre de Bruxelles (ULB), Bruxelles, Belgium
\\
\bf{3} UMR EME, Institut de Recherche pour le D\'eveloppement (IRD), Victoria, Seychelles
\\
$\ast$ E-mail: manuela.capello@ird.fr
\end{flushleft}

\section*{Abstract}

Demonstrating and quantifying the respective roles of social interactions and external stimuli governing fish dynamics is key to understanding fish spatial distribution. 
If seminal studies have contributed to our understanding of fish spatial organization in schools, little experimental information is available on fish in their natural 
environment, where aggregations often occur in the presence of spatial heterogeneities. 
Here, we applied novel modeling approaches coupled to accurate acoustic tracking for studying the dynamics of a group of gregarious fish in a heterogeneous environment. 
To this purpose, we acoustically tracked with submeter resolution the positions of twelve small pelagic fish ({\it Selar crumenophthalmus}) in the presence of an anchored floating object, constituting a point of attraction for several fish species. We constructed a field-based model for aggregated-fish dynamics, deriving effective interactions for both social and external stimuli from experiments. We tuned the model parameters that best fit the experimental data and quantified the importance of social interactions in the aggregation, providing an explanation for the spatial structure of fish aggregations found around floating objects. Our results can be generalized to other gregarious species and contexts as long as it is possible to observe the fine-scale movements of a subset of individuals.


\section*{Introduction}
Despite the social and economic importance of fisheries, quantitative tools capable of predicting fish 
distribution and its variations with respect to environmental changes and human activities are still missing.
{The main approaches that are currently used in fisheries management require going beyond the study of isolated target fish 
species and demand taking into account intra- and inter-specific interactions, behavioral factors as well as responses to the environment \cite{BN1,BN2}. 
More generally, they raise fundamental questions on animal organization in a natural environment.
}
Demonstrating and quantifying the respective roles of external factors and social influences is key to understanding the spatial distribution and organization of animals in their environment. 
In the last decade, several studies have shown how social interactions govern the dynamics of animal groups, such as fish schools, bird flocks, sheep herds or aggregations of insects \cite{BN3,BN4,BN5,BN6-2,BN6bis,BN6-3,BN6}. However, in a natural environment, animal aggregations often occur in the presence of environmental heterogeneities, constituting a point of attraction for feeding, sheltering or other behaviors. 
This demands the creation of dedicated analytical and modeling tools capable of taking into account interactions,
both with the other individuals and with a heterogeneous environment. 
There is substantial evidence that social behavior is important for many fish species,
 yet within the existing experimental and modeling approaches, it is difficult to quantify the respective roles played by fish social interactions and external stimuli in their spatial distributions. On one hand, microscale models \cite{BN6,BN7,BN7-2,BN7-3,BN7-4,BN7-5}, which consider individuals embedded in an homogeneous environment, can explain the observed schooling and milling phenomena but cannot be used to make predictions on the spatial distribution of the different fish species due to the difficulty in estimating model parameters experimentally. On the other hand, macroscoscale models \cite{BN8} capable of incorporating the response of fish populations to environmental gradients for different species do not take into account behavioral features that could play a crucial role in the fish spatial distribution. 
In this study, we worked at an intermediate scale, 
deriving a field-based model for fish dynamics that could incorporate 
at the same time the basic ingredients for fish response to social stimuli and environmental heterogeneities.
Remarkably, this modeling approach can be applied to a large variety of phenomena whenever the spatial distribution 
of individuals results from the mutual response to environmental and social interactions. 
We used the case of a group of fish in the presence of a floating object, known in the literature as a Fish Aggregation Device (FAD) \cite{BN9,BN10}. FADs can be artificial or natural floating structures, either drifting or anchored. They have been massively deployed by commercial fisheries since the eighties because they constitute a point of attraction for many fish species. However, the reason fish aggregate around FADs is still unknown. 
Understanding aggregated-fish behavior is becoming more urgent, due to the large and continually growing exploitation of FADs. Recently, concerns that FADs may act 
as ecological traps for fish have been voiced \cite{BN11}, {suggesting that the retaining character of FADs may alter the biological characteristics of fish populations associated with them, like migration, growth, condition factors, predation and natural mortality.} 
The validity of these scenarios strongly depends on the type of mechanisms leading to aggregation, and this requires further investigation. To this end, precise information on the range and structure of fish aggregations is needed. To date, several acoustic surveys have been done in order to characterize 
fish aggregations around FADs \cite{BN12,BN13,BN14}. 
However, no information was available so far on aggregated-fish dynamics at submeter scales, where fish behavior could be studied in detail.

In this paper, we introduced a new modeling approach, coupled to accurate acoustic tracking measurements, capable of quantifying 
the driving forces leading to aggregation.
We considered an obligate schooling small pelagic fish species, {\it Selar crumenophthalmus}, in the proximity of a FAD located in Saint Paul's Bay at Reunion Island (West Indian Ocean) \cite{BN15}. The FAD was a 12-m boat fixed at 17 m depth by five anchors to prevent any movement.
Twelve fish were tagged with HTI\texttrademark acoustic pingers (Hydroacoustic Technology Inc., Seattle, USA) and released next to the FAD, along with other non-tagged individuals. The 3D tracking of each tagged fish (one position every second with sub-meter resolution) was possible within a radius of approximately 50 m from the FAD with the use of an HTI\texttrademark acoustic detection system. Experimental data collected during one hour were used to construct a model for aggregated-fish dynamics, which took into account the possible interactions with the FAD and the other tagged fish, while all other factors (e.g., light, food abundance, and currents) were considered constant during this period of observation. Model parameters were fine-tuned with experimental data, which allowed us to quantify the interplay between social interactions and attraction to the FAD and to gain insights on the aggregation phenomena.

\section*{Results}

\subsection*{Experimental data analysis}
\begin{figure}[!ht]
\begin{center}
\centerline{\includegraphics[width=8.7cm]{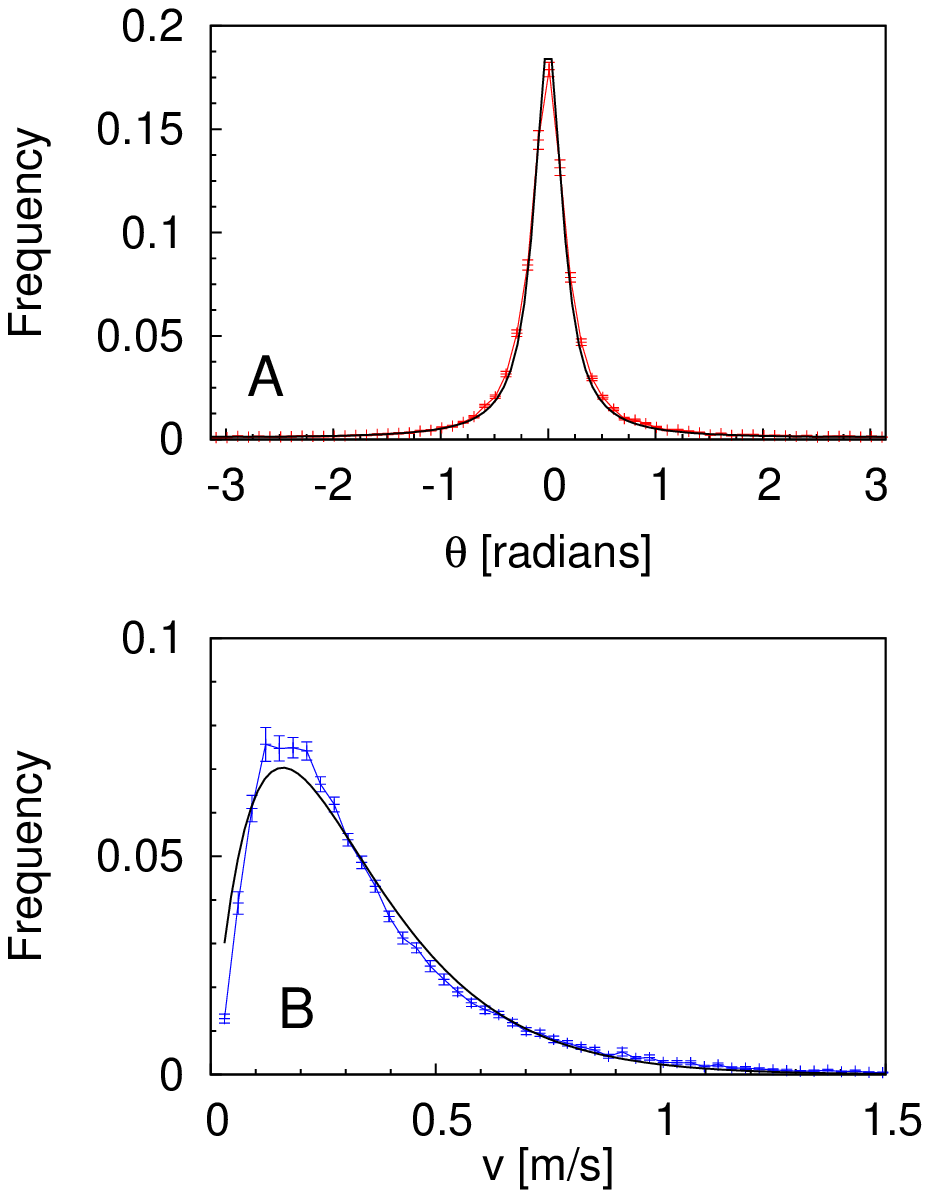}}
\caption{{\bf Analysis of individual-fish dynamics} (A) Turning angle distribution: experimental points (red) and fit (black) with the Wrapped Cauchy Distribution$\Phi(\theta)=\frac{1}{2\pi}\frac{\sinh(\rho)}
{\cosh(\rho)-\cos(\theta)}$ with fitting parameter $\rho=0.16$.
(B) Individual swimming speed distribution: experimental points (blue) and fit (black) with the Gamma distribution
$f(x)=\, c\,  x\, \exp(-x/s)$
with scale parameter $s=0.16$ and normalization constant $c=1.2$.
}\label{fig1}
\end{center}
\end{figure}

\begin{figure}[!ht]
\begin{center}
\centerline{\includegraphics[width=8.7cm]{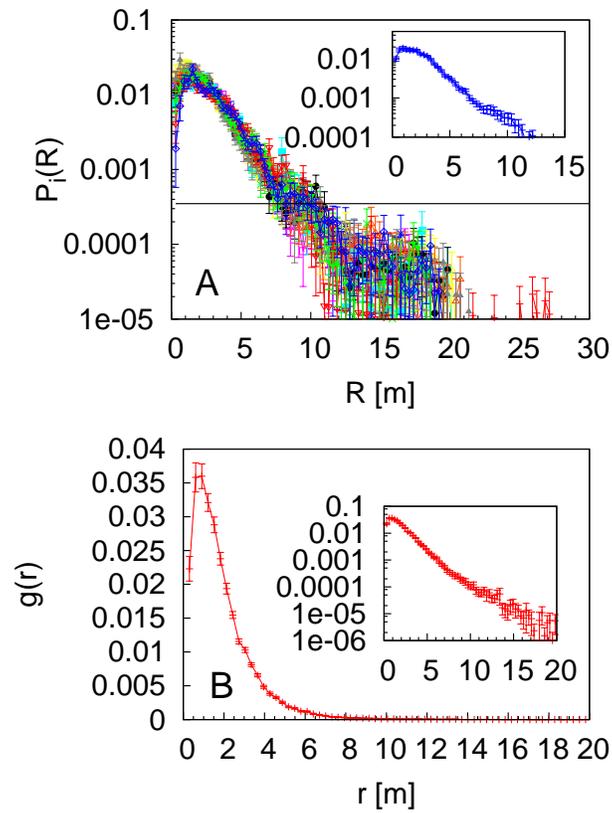}}
\caption{{\bf Analysis of fish spatial distribution around the FAD}:
(A) Radial distribution around the FAD for each of the 12 tagged fish (represented by different colors) in semilogarithmic scale.
The horizontal line indicates the behavior of $P(R)$
for a fish having a homogeneous distribution in a circle of radius equal to 30 m. Inset: mean value over all fish.
(B) Fish pair-correlation function. Inset: the same in semilogarithmic scale.
}\label{fig2}
\end{center}
\end{figure}

Two variables were considered to characterize the individual fish dynamics: turning angle and swimming speed 
in the $xy$ plane.  The turning-angle distribution was well described by a wrapped Cauchy distribution, with a sharp peak centered at zero (Fig.\ref{fig1}A). At the time scale of our observations, no evidence of correlation among subsequent turning angles was found \cite{BN17}. The speed distribution had a maximum at approximately
$0.16\pm 0.03$ m/s (Fig.\ref{fig1}B).
Based on the observation that the average fork length of our tagged fish was $0.17\pm 0.02$ m, this finding was compatible with the widely accepted kinetic rule of 1 body-length/second. At higher speeds, the distribution decayed exponentially.
The recorded fish trajectories in the $xy$ plane (see Fig.S2 in Supplementary material) demonstrated a radial symmetry around the FAD.
Therefore, in order to analyze the fish spatial distribution, we calculated the time-averaged {\it radial distribution} $P_i(R)$ for each fish $i$ with respect to the FAD position,
where $R$ is the radial distance from the FAD in the $xy$ plane (Fig.\ref{fig2}A).
This quantity gave information on the probability to find a fish at a distance $R$ from the FAD \cite{BN35}.
Remarkably, all of the tagged fish showed the same radial distribution.
Close to the FAD, there was a region (at a distance smaller than 2 m) characterized by a high and constant $P_i(R)$. Thereafter, $P_i(R)$ decreased exponentially up to
a distance of approximately 10 m, where the radial distribution was of the order of the constant distribution associated to a fish occupying homogeneously the detection area.
This scale sets the boundary of the zone of aggregation. The interactions between tagged fish, as well as their variations in space,
were investigated through the time-averaged fish {\it pair-correlation function} $g(r)$ \cite{BN35}, where $r$ is the radial distance among fish pairs (Fig.\ref{fig2}B).
The depletion of $g(r)$ for distances smaller than 0.3 m indicated a zone of repulsion. At intermediate distances, the fish pair-correlation was constant and maximum, revealing a zone of comfort. For distances larger than 0.9 m, the pair-correlation decayed exponentially, revealing that it was less probable to find inter-individual distances in this range.
Both the speed and turning angle distribution, as well as the pair-correlation function, were independent of the radial distance from the FAD, 
see Fig.S4 in Supplementary material. 

\subsection*{Model results}
In order to gain quantitative insights into the fish response to both the FAD and the other fish,
we derived effective interactions from the experimental quantities discussed above.
Expressing the average fish-radial distribution $P(R)$ around the FAD as the exponential of a Boltzmann weight $P(R)\sim \exp[-V_{FAD}(R)]$ \cite{BN18},
we obtained the effective fish-FAD interaction $V_{FAD}(R)\sim –\log[P(R)]$. From the behavior of the experimental $P(R)$ this lead to the following expression for the fish-FAD interaction:
\begin{equation}\label{eq1}
V_{FAD} (R)= 
\left\{ \begin{array}{ll}
  \mbox{const} & \mbox{for } R \leq R_{st} \\
  \alpha R     & \mbox{for } R > R_{st} 
 \end{array}\right.
\end{equation}
where $R_{st}$ is the stationarity radius, found at 2 m, which corresponds to the region with constant probability to find the fish.
Beyond this distance was the zone of FAD attraction, where fish responded to the presence of the FAD through a constant attractive force,
whose strength, $\alpha$, was obtained by comparison with experimental data, as shown below.
This zone was assumed to be much larger than the other spatial scales governing fish dynamics \cite{BN19}. The definition of its boundaries went beyond
the scope and experimental limits of this study. 

\begin{figure}[!ht]
\begin{center}
\includegraphics[width=12.35cm]{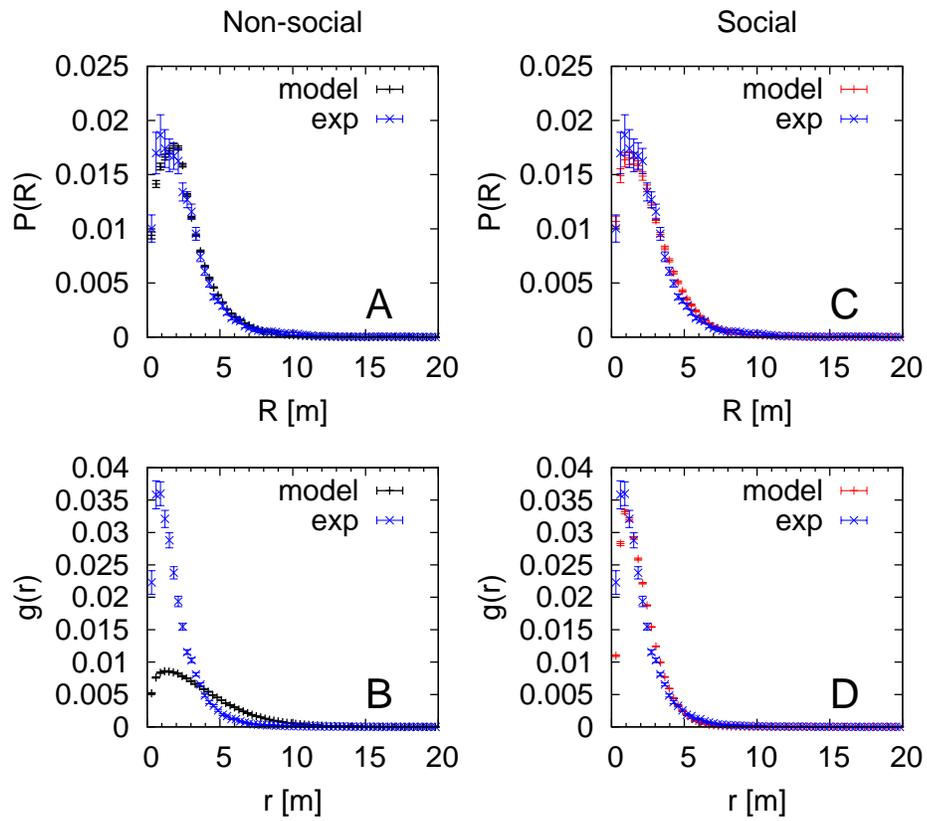}
\caption{
{\bf Comparison among optimized
model and experimental results for the fish radial distribution $P(R)$ around the FAD and the pair correlation function $g(r)$}.
Left panels: non-social fish model (A) radial distribution and (B) pair-correlation function.
Right panels: social fish model (C) radial distribution and (D) pair-correlation function.
}\label{fig3}
\end{center}
\end{figure}

\begin{figure}[!ht]
\begin{center}
\includegraphics[width=8.7cm]{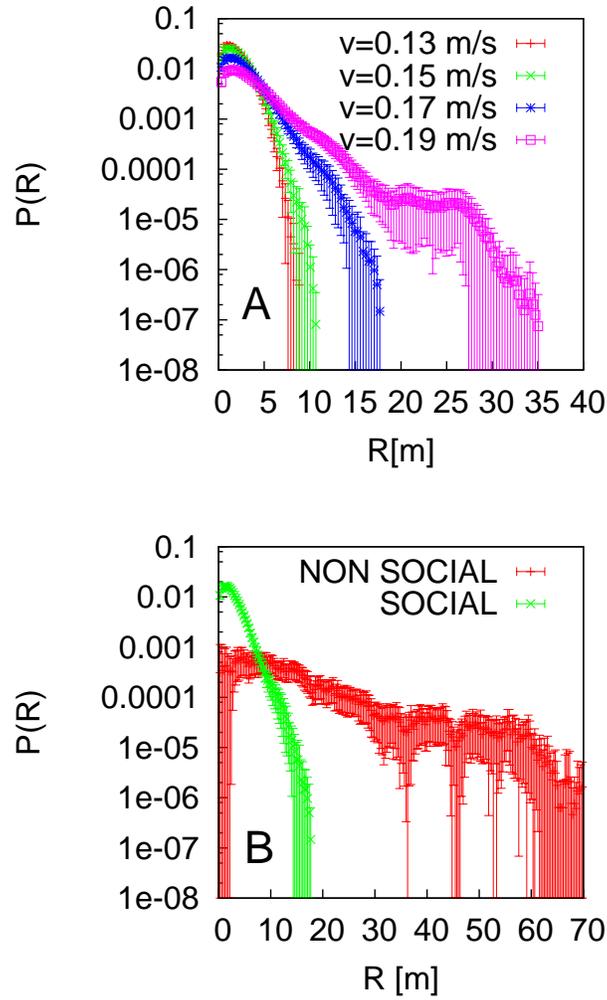}
\caption{{\bf Role of individual swimming speed and social interaction on fish aggregation.}
Radial distribution around the FAD obtained from the optimized model (with fish-fish interactions) when varying
(A) the individual fish swimming speed $v$,
(B) the social interaction parameter $\beta$ (NON-SOCIAL corresponds to $\beta=0$ and SOCIAL indicates the optimized model).
}\label{fig4}
\end{center}
\end{figure}

\begin{figure}[!ht]
\begin{center}
\includegraphics[width=8.7cm]{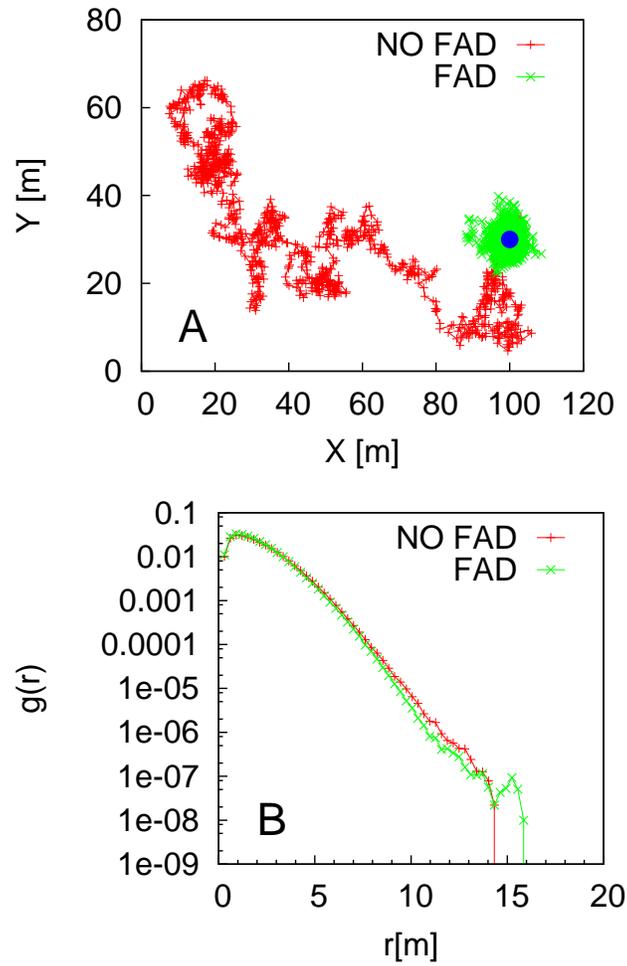}
\caption{{\bf Role of FAD attraction on fish aggregation.} Optimized model prediction for the (A) fish-group baricenter (FAD represented with a blue circle) and (B) pair-correlation in the presence/absence of a FAD.
}\label{fig5}
\end{center}
\end{figure}

Analogously, we derived effective fish-fish interactions from the pair-correlation function, defining  $V_{fish}(r)\sim –\log[g(r)]$
\cite{BN18}.
According to the behavior of $g(r)$, we identified three main zones for the fish-fish interaction: a zone of repulsion within a radius
$r_{rep}=$0.3 m, a zone of comfort up to $r_{comf}=$0.9 m and a zone of attraction at larger distances. This lead to the following effective fish-fish interaction:
\begin{equation}\label{eq2}
V_{fish} (r)= 
\left\{ \begin{array}{ll}
  -v\;r        & \mbox{for } r \leq r_{rep} \\ 
  \mbox{const} & \mbox{for } r_{rep}<r \leq r_{comf} \\
  \beta r     & \mbox{for }  r>r_{comf}
 \end{array}\right.,
\end{equation}
where the analytic form of the short-range repulsion was assumed linear for simplicity,
and $v$ is the individual fish speed (v=0.17 m/s, following the rule of 1 body-length/second). The parameter, $\beta$,
setting the degree of attraction between fish, was obtained by comparison with experimental data, as shown below.
Given these effective interactions, we modeled the system following a correlated random walk dynamics (CRW) embedded in a force field
\cite{BN20,BN21}. In order to evaluate the role of the fish-FAD interaction in the measured quantities, we first compared results from the non-social model
(i.e., no fish-fish interaction) with experiments.
Taking the expression of the effective fish-FAD interaction in Eq.\ref{eq1}, the only free parameter
 was the FAD attraction strength, $\alpha$.
We estimated the value of $\alpha$ that minimized the sum of squared residuals (SSR) for the radial fish distribution $P(R)$ averaged over all fish (Fig.\ref{fig3}A).The pair-correlation function (Fig.\ref{fig3}B) calculated for the optimized non-social system was different than the one found from the experimental data,
signaling that the observed fish aggregation around the FAD was not purely a consequence of the FAD attraction.
We then adjusted the parameters $\alpha$ and $\beta$ in Eq.\ref{eq1} and Eq.\ref{eq2} to find the minimum of the SSR for both the fish radial distribution and the pair-correlation function (see Table 1). Indeed, adding fish-fish interactions resulted in agreement between the model and experimental 
data for all quantities and showed that the fish dynamics around the FAD was also the fingerprint of a true fish-fish interaction (Fig.\ref{fig3}C and \ref{fig3}D).
Moreover, starting from the optimized model that best fit the experimental data, we studied the system sensitivity to changes of one model parameter at a time. First, we calculated the radial distribution around the FAD for different values of the individual fish speed, $v$,
keeping constant the fish-fish and fish-FAD interactions. Small changes in $v$ affected the zone of 
aggregation significantly, with an aggregation radius increasing with speed (Fig.\ref{fig4}A). Next, we analyzed the role of social interactions, keeping constant the FAD attraction. We obtained that 
non-social fish should have a larger dispersion around the FAD, with an aggregation radius of about 
60 m rather than 10 m for social fish (Fig.\ref{fig4}B). Finally, we studied the fish-group dynamics
 in the absence of a FAD. The fish-group baricenter performed a random walk and explored the environment (Fig.\ref{fig5}A), with the group staying compact.
This is clear from the comparison of the fish pair-correlation in the presence/absence of a FAD (Fig.\ref{fig5}B).

\section*{Discussion}
Experimental data revealed the existence of a sharp zone of aggregation, where {\it Selar crumenophthalmus} concentrate at small distances
from the FAD. All tagged fish exhibited the same behavior and mostly stayed within 10 m from the FAD.
We could distinguish a stationarity region very close to the FAD ($<$ 2 m), with a constant and high probability of finding fish, as well as a larger-ranged zone (2-10 m) with an exponentially decaying probability. This level of spatial accuracy, as well as information on the specific shape of the zone of aggregation, cannot be reached with standard acoustic techniques \cite{BN12,BN13,BN14}, 
revealing that this experimental approach is ideally suited for understanding the behavior of fish aggregated to FADs at a fine scale.
In order to identify the factors that shaped the zone of aggregation, we constructed the simplest model of fish dynamics that would take the main ingredients into consideration, where the fish-FAD and the fish-fish interaction was deduced from the spatial distribution of the tagged fish. Although the attraction of the FAD alone allowed modeling of the fish aggregation to the object, the matching between experimental and modeled data was only possible when taking into account the fish-fish interaction. The optimized value of the FAD attraction, $\alpha$, was much smaller than the CRW individual speed parameter, $v$, indicating that the stochastic term was playing a major role. In other words, when fish stayed in the zone of aggregation, their movements were dominated by the correlated-random walk component.
The optimized value of the fish-fish attraction, $\beta$, was larger than $\alpha$ and was necessary in order to reproduce the experimental pair-correlation function. 
In this way, the model highlighted the important role of social interactions in the distribution of fish around a FAD. Indeed, although we only tagged some individuals from a group, the time average of our tagged-fish pair-correlation offered insights into the entire fish aggregation near the FAD. The effective fish-fish interaction implicitly took into account the presence of other non-tagged fish in the system. This represents a key improvement in the study of social behavior of wild animals in their environment, as an exhaustive observation of all members of a group is almost never possible. Moreover, our optimized model shows that the fish-fish interaction alone, estimated through our field-based approach near the FAD, can ensure a stable group dynamic, even in the absence of a FAD. This prediction signals that the observed aggregation is a stable entity and could be the precursor of schooling \cite{BN10,BN15}.

Finally, our model allowed us to make predictions for different values of the parameters controlling the aggregated-fish dynamics. Within our scheme, the boundaries of the zone of aggregation appear to be very sensitive to both changes in the fish speed and social interaction. Indeed, social individuals would be closer to the FAD than non-social fish, implying an amplification of the individual response to the FAD attraction in a social group \cite{BN22}. Moreover, fish characterized by a high speed would have a wider zone of aggregation, signaling possible correlations among the fish swimming speed (or size) and their spatial distribution around FADs. This provides a unique explanation for the structure of fish aggregations around floating objects, where smaller species (or smaller individuals), characterized by a smaller swimming speed, are found closer to FADs than larger fish \cite{BN12,BN13,BN14}. This approach can be used to study the fish dynamics of other species and predict the occurrence of different shells of fish concentrations around the FAD. A fine-tuned analysis of the strength of interactions would be required to better assess the aggregation radius of each species. However, by fixing the same ratio among all quantities and simply scaling all the parameters governing the dynamics of a factor, F, our model predicts lower and upper bounds for the zone of aggregation of social and non-social fish  at 10xF m and 60xF m, respectively. By using a value of 5 for F, a scale that could correspond to the individual swimming speed and size of a tuna, would result in an aggregation radius between 50m and 300m. This lower bound is close to the center of mass for tuna aggregations found in recent acoustic experiments \cite{BN13,BN14}, suggesting a potentially strong social effect for tuna aggregated to FADs.
\section*{Materials and Methods}
\subsection*{Ethic Statement}
The fish experimental protocols were permitted under the Aquarium of Reunion Island animal care certificates delivered by the French Veterinary Medicine Directorate. Protocols were carried out with the authority of the National Veterinary School of Nantes (France) validating a certificate 
of training in animal experimentation and a degree in experimental surgery on fish.
\subsection*{Experimental setting}
The experiment was conducted in open field of a shallow-water region (17 m depth in average) in the center of Saint Paul's Bay in Reunion Island (South Western Indian Ocean). The HTI\texttrademark Acoustic Tag Tracking System (Model 290) was composed of five hydrophones connected by cables to the Acoustic Tag Receivers system embedded on the boat. These hydrophones surrounded the boat in a square of approximately 100 meters per side (see Fig.S1 in Supplementary Information). The hydrophones were arranged at the surface and near the sea bed to allow for optimal reception. The cables connecting the hydrophones to the boat reached the sea bottom straight below the boat, constituting a vertical submerged structure whose position was taken as our FAD position (Fig.S2 in Supplementary Information). The HTI\texttrademark acoustic tags (Model 795) were 7 mm diameter, 17 mm length and 1.5 g weight in the water. This weight was less than 0.1\% of the mean fish weight.
We were therefore confident that the tags did not affect the buoyancy of the fish {\cite{buoyancy}}. The in situ test led us to choose a pulse duration of 4 msec. In order to discriminate fish, the repetition rate (number of transmissions per second) was programmed to be different for each tagged fish and ranged between 1.43 and 1.16 $s^{-1}$.
These settings were optimal for the duration of our experience, with a theoretical period of life tags of six days. Data processing involved two steps. First, the acoustic record of each tag on each of the five hydrophones was manually proofed using HTI\texttrademark Mark Tags Software to exclude acoustic noise.
Second, files were processed in HTI\texttrademark Acoustic Tag program to track acoustic echoes.
This procedure used a hyperbolic algorithm to solve for the transmitter 3D position.
In addition, a time-stamp was calculated so that the transmitter was referenced in both space and time. The accuracy of the position in the horizontal plane ranged from 0.1 to 0.3 m in the monitoring network. In the vertical direction, the accuracy was lower (around 1.0 m). We measured current through the Aanderaa RMC 9 Self Recording Current Meter, which was fixed under the boat (at 5 meters depth) in order to record the horizontal current speed and direction. The minimum-current period recorded by the current meter occurred when the direction of the tidal current reversed (Fig.S3 in Supplementary Information), between 13:00 and 14:00.
\subsection*{Fish Species and Tagging procedure}
The fish species we studied was the big-eye scad ({\it Selar crumenophthalmus}). It is a small coastal pelagic fish common in the circumtropical area \cite{BN31} and is an obligate schooler (i.e., unable to survive outside a fish school \cite{BN32,BN33}), {which is known to associate around FADs \cite{taquet}.
In Reunion Island,
Saint Paul's bay is the main area where this species is caught. Here, traditional beach seiners target shoals of bigeye scads aggregated around anchored FADs near the shore.
} 
Forty fish were caught using hand lines {in Saint Paul's bay}, transported in baskets and maintained in tanks at the Aquarium of Reunion Island during ten days for acclimation. They were fed and treated with a solution of methylene blue (from the first day) and copper sulfate to kill bacteria and to prevent the proliferation of fungi. Most fish showed only superficial wounds, which were caused by fishing (hook) or handling. The tagging operation was carried out on the 1st of May 2003. Twelve fish were anesthetized with a solution of clove oil \cite{BN34}. The acoustic tags were implanted gastrically by ingurgitation, 
{a tagging technique well suited for short-term experiments \cite{martinelli}}. 
The fish were held for two subsequent days in tanks to ensure fish survival and tag retention. No further mortality was observed in either tagged or untagged fish during this period. All fish (tagged and non-tagged) were released on the 3rd of May 2003 at 12:00 in the proximity of the boat, {anchored in the nearby of the fishing location}. Based on visual observation, we could estimate that fish immediately formed a small school. All fish stayed within the zone of detection until 19:00, with very few excursions outside the range of detection. Three fish stayed at night, leaving the zone the subsequent morning, while the others left around 19:00. Four fish made short visits during the second and third day of the experiment.

\subsection*{Methods for data analysis}
Experimental data analysis concentrated on one hour, between 13:00 and 14:00 of the first day, when the current was negligible and all of the tagged fish were present. This allowed us to collect good statistics, with about 3600 positions for each of the twelve tagged fish. During the rest of the day, our results still held, but we observed a shift in the position of the fish baricenter, due to non-negligible
current effects.
Due to the system geometry, where the floating object was associated to a submerged vertical structure reaching the sea-bottom (see Fig.S1), 
data analysis focused on the $xy$ plane, integrating over the vertical direction. 
This approach was supported by previous acoustic survey measurements \cite{BN12,BN13,BN14} 
were the fish spatial distribution along $z$ was not affected by the presence of the FAD 
but rather depended on the fish species. 
The presence of each fish $i$ around the FAD was analyzed through the time-averaged radial distribution function:
\[P_i(R)=\left\langle \frac{\delta(R-R_i(t))}{2\pi R \; dR}\right\rangle\]where the delta function selects the fish positions $R_i(t)$ at time $t$ with radial distance from the FAD within the interval $[R-dR, 
R]$ \cite{BN36},
and brackets denote the time average. The denominator corresponds to the area of the ring of radius $R$ and width $dR$ around a FAD.
With this denominator, the quantity $P_i(R)$ was normalized:
\[\int P_i(R) 2\pi R dR=1\]
Therefore, $P_i(R)$ could be interpreted as the probability of the presence of fish $i$ at distance $R$ from the FAD, per unit area.The strength of our approach resided in the high number of sampled points, which clearly allowed us to speak in terms of probabilities.
 We chose $dR=0.3$m,
which was compatible with the experimental precision for fish detection in the $xy$ plane.
Because we had detailed spatial information concerning several fish, we calculated the fish pair-correlation function (or pair-distribution function)
$g(r)$ among synchronous fish \cite{BN35}. This gave us an understanding of fish interactions and their variations in space. We took synchronicity intervals of 1 second because this time frame was sufficient to obtain a large number of synchronous fish. We calculated the time-averaged fish pair-correlation function $g(r)$, which in two dimensions can be written as:
\[g(r)=\left\langle \frac{1}{N(t)} \frac{\sum_{ij}\delta(r-r_{ij}(t))}
{2\pi rdr}
\right\rangle\]
where $N(t)$ is the number of coplanar pairs detected in the temporal interval $[t,t+1s]$ and $r_{ij}(t)$ is the planar distance among synchronous
 fish $i$ and $j$.
Coplanarity was established when two fish were within 1 m in the $z$-direction, which was compatible with our experimental accuracy in the vertical direction.
The delta function selects fish pairs at planar distance in the range $[r-dr,r]$, with $dr=$0.3m, and brackets denote the time average.

\subsection*{Model definition}
Lagrangian dynamics \cite{BN36,BN37} are characterized by the use of stochastic differential equations that describe the evolution of the positions of each individual in time. The system evolves under the effect of both deterministic forces and a random component. Here, we used a variant of the Lagrangian dynamics, the correlated random walk (CRW) \cite{BN20,BN21}, in the presence of a force field. In the CRW model, an animal makes discrete steps, with turning angles sampled from a given probability distribution. At each step, the turning angle is independent of the previous one. Here, in addition to isolated-fish CRW dynamics, fish movement was influenced deterministically.
The time evolution of the position of fish $i$ in the plane followed the equations:
\begin{eqnarray}\label{eq3} 
x_i(t+\Delta t)&=&x_i(t)+v\cos(\omega_t+\theta)\Delta t +\Delta t F_x(t) \nonumber \\
y_i(t+\Delta t)&=&y_i(t)+v\sin(\omega_t+\theta)\Delta t +\Delta t F_y(t)
\end{eqnarray}
where $x_i(t)$ and $y_i(t)$ are the $x$ and $y$ component of the position of fish $i$ at time $t$, and $\Delta t$ was the time step.
The first terms on the right hand side constituted the standard CRW dynamics, with $v$ being a constant defining the individual swimming speed.
Concerning the angular component, $\omega_t$ corresponded to the fish orientation angle in our reference frame at time $t$ and $\theta$ was a random number
taken from a probability distribution $\Phi(\theta)$ that sets the turning angle. This probability distribution, as well as the
constant for the individual swimming speed, was taken from experiments.
In particular, $v$ corresponded to the standard rule of 1 body length per second and $\Phi(\theta)$ followed a wrapped Cauchy distribution.
The last terms in Eq.\ref{eq3} constituted the deterministic part of the fish dynamics, where $F_x(t)= - \frac{dV(\{x_j,y_j)\}}{dx_i}$
 $\left(F_y= - \frac{dV(\{x_j,y_j\})}{dy_i}\right)$
was the $x$ $(y)$ component of the deterministic force associated with a potential $V(\{x_j,y_j\})$ at time $t$, depending on fish positions.
We considered additivity in the forces. Our potential had the form:
\[V({x_i,y_i})=V_{FAD}(R_i)+\sum_{ij}V_{fish}(r_{ij})\]
where the first term was the FAD potential, and the second term set the fish-fish interaction,
with $R_i=\sqrt{(x_{FAD}-x_i)^2+(y_{FAD}-y_i)^2}$ being the radial distance among the FAD and fish $i$ and $r_{ij}=\sqrt{(x_{j}-x_i)^2+(y_{j}-y_i)^2}$ being the distance among fish $i$
and fish $j$.
The analytic forms of these potentials were derived from the experimental radial distribution $P(R)$ and the pair correlation function $g(r)$
\cite{BN40}. The optimized model parameters are shown in Table 1.

\section*{Acknowledgments}
We thank G. Potin, M. Taquet, L. Bigot, P. Durville, M. Timko, G. Fritsh for their help in fishing and tagging operations
and P. Fr\'eon, F. Becca, O. Zagordi and G. Nicolis for interesting discussions. 

\bibliography{mcapello-et-al}

\newpage

\section*{Figures}

\newpage

\newpage

\newpage

\newpage

\newpage

\section*{Tables}
\begin{table}[!ht]
\caption{
\bf{Model parameters}}
\begin{center}
  \begin{tabular}{| l | c | c | c | }
    \hline
Model Parameter                          & Symbol    & Non-social & Social\\ \hline \hline
Number of fish                           & $N$       & 12              & 12         \\ \hline
Individual speed [m/s]                   & $v$       & 0.17            &0.17        \\ \hline
Wrapped Cauchy distribution parameter    & $\rho$    &0.16             &0.16        \\ \hline
FAD Stationarity radius [m]              & $R_{st}$  &2                &2               \\ \hline    
FAD attracting potential strength *      & $\alpha$  &0.014            &0.0015      \\ \hline
Fish-fish repulsion radius [m]           & $r_{rep}$ & --              &0.3     \\ \hline
Fish-fish comfort radius [m]             & $r_{conf}$& --              &0.9             \\ \hline
Fish-fish attracting potential strength * & $\beta$  & --              &0.003    \\\hline
    \hline
  \end{tabular}
\end{center}
\begin{flushleft}
{Model parameters used to fit the experimental data in Fig.3.}
Stars indicate the free parameters estimated through minimization of the SSR
on the radial distribution function and the pair correlation function.
Third column, non-social fish model. Forth column, social fish model.
\end{flushleft}
\end{table}

\newpage

\section*{Supporting Information }
\renewcommand{\thefigure}{S\arabic{figure}}
\setcounter{figure}{0}

\begin{figure}[!ht]
\begin{center}
\includegraphics[width=8.7cm]{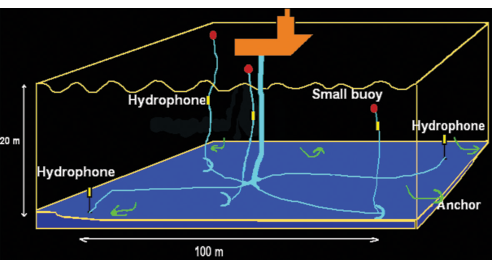}
\caption{{The HTI\texttrademark experimental setting.} The boat, with the cables underneath, represents the floating object or 'FAD'.}
\end{center}
\end{figure}

\begin{figure}[!ht]
\begin{center}
\includegraphics[width=8.7cm]{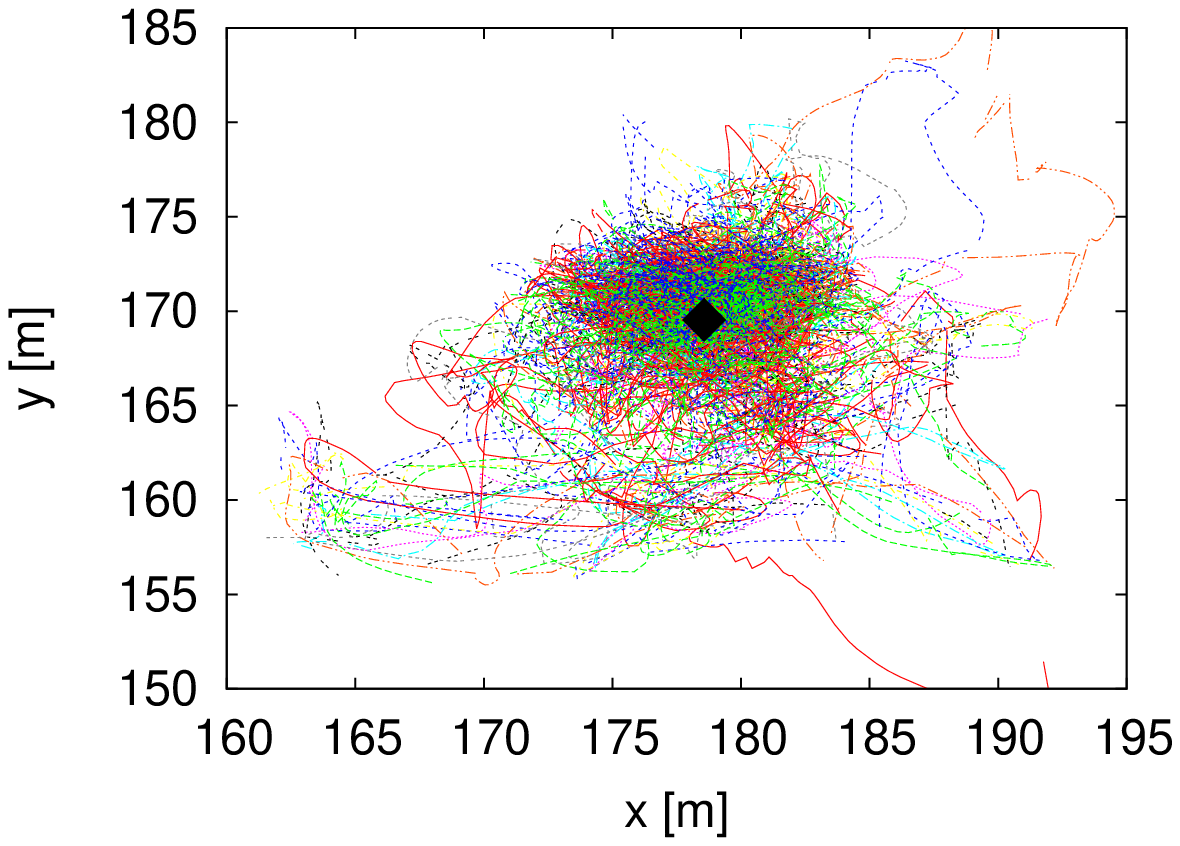}
\caption{{Trajectories of the tracked fish in the $xy$ plane around the FAD from 13:00 to 14:00} 
Different colors indicate different fish and black point indicates the FAD position.}
\end{center}
\end{figure}

\begin{figure}[!ht]
\begin{center}
\includegraphics[width=8.7cm]{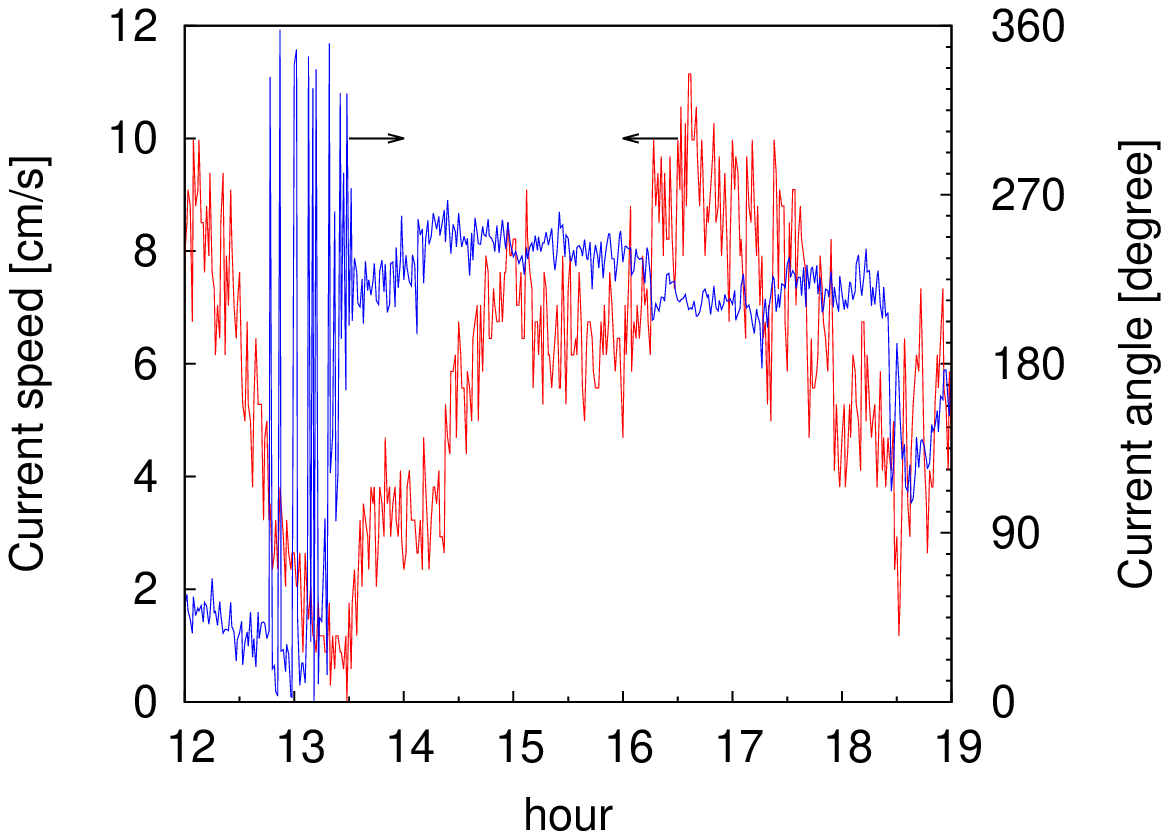}
\caption{Current speed (red line) in cm/s and current angle (blue line), with respect to the North (East=90; West=270) recorded during the experiment.}
\end{center}
\end{figure}

\begin{figure}[!ht]
\begin{center}
\includegraphics[width=8.7cm]{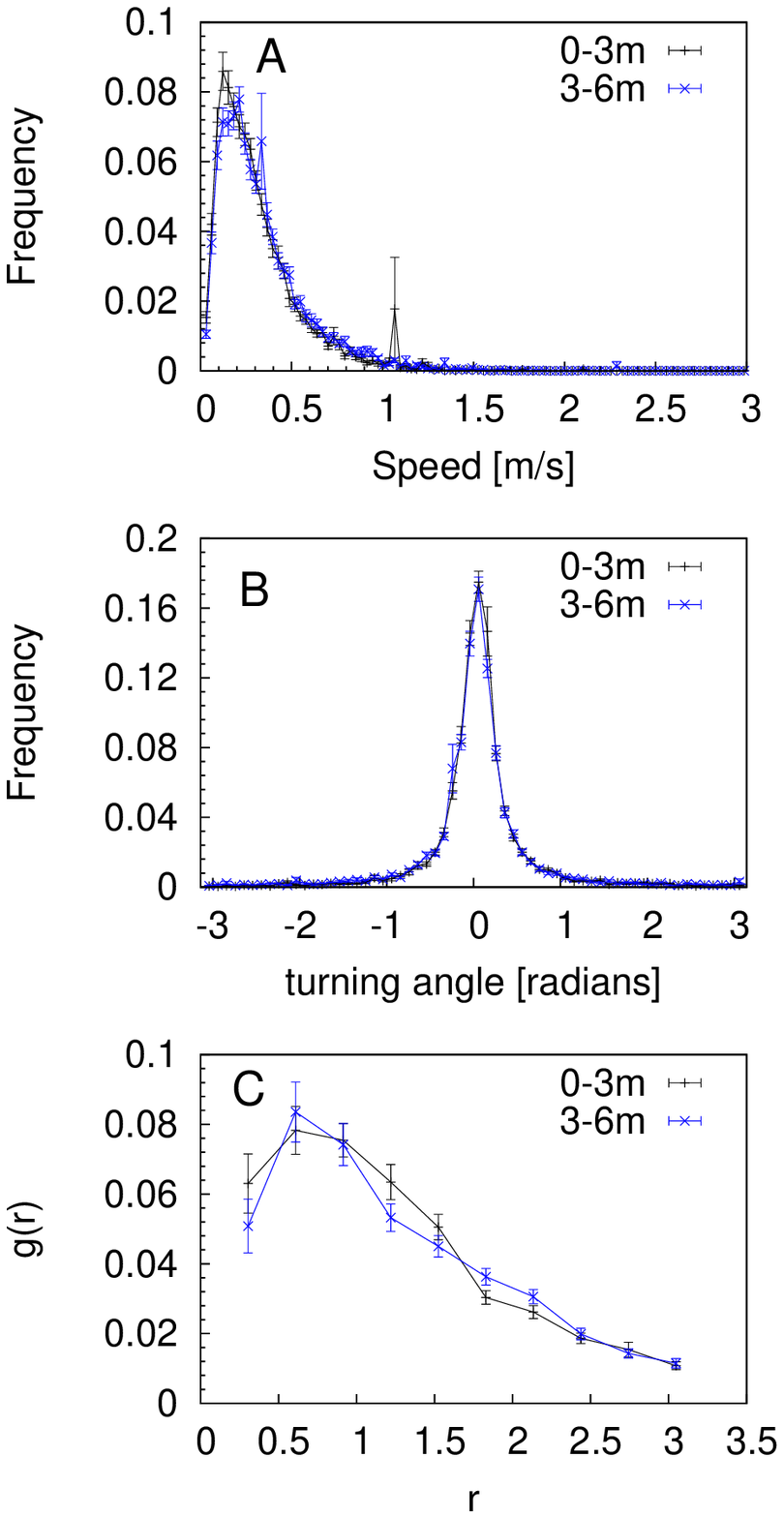}
\caption{Swimming speed distribution (A), turning angle distribution (B) and pair-correlation function (C), calculated at different radial distances from the FAD.
}
\end{center}
\end{figure}

\end{document}